\documentclass{article}

\usepackage{arxiv}

\usepackage[utf8]{inputenc}
\usepackage[T1]{fontenc}
\usepackage{hyperref}
\usepackage{url}
\usepackage{booktabs}
\usepackage{amsfonts}
\usepackage{nicefrac}
\usepackage{microtype}
\usepackage{cleveref}
\usepackage{graphicx}
\usepackage{natbib}
\usepackage{doi}
\usepackage{enumitem}
\usepackage{subcaption}
\usepackage{xcolor}
\usepackage{tikz}
\usepackage{listings}
\graphicspath{{figures/}}
\usetikzlibrary{shapes, arrows.meta, positioning, fit, backgrounds}

\title{ECNUClaw: A Learner-Profiled Intelligent Study Companion \\ Framework for K-12 Personalized Education}

\date{}

\usepackage{authblk}

\author[1,2]{Yizhou Zhou}
\author[1,3]{Jiayin Li}
\author[1,2]{Zhi Zhang\thanks{Corresponding author. Email: \texttt{zhangzhi@ai.ecnu.edu.cn}}}

\affil[1]{Shanghai Institute of AI Education, East China Normal University, Shanghai 200062, China}
\affil[2]{AI Education Laboratory, East China Normal University, Shanghai 200062, China}
\affil[3]{Institute of Higher Education, Faculty of Education, East China Normal University, Shanghai 200062, China}

\renewcommand{\shorttitle}{\textit{ECNUClaw: Learner-Profiled Study Companion}}

\hypersetup{
pdftitle={ECNUClaw: A Learner-Profiled Intelligent Study Companion Framework for K-12 Personalized Education},
pdfsubject={cs.AI, cs.CL, cs.CY},
pdfauthor={Yizhou Zhou, Jiayin Li, Zhi Zhang},
pdfkeywords={Learner Profiling, Intelligent Tutoring Systems, K-12 Education, Personalized Learning, Education Brain, Agent Framework},
}

\begin{document}
\maketitle

\begin{abstract}
We introduce ECNUClaw, an open-source framework for building learner-profiled intelligent study companions in K-12 education. The system constructs and maintains a five-dimension learner profile---covering cognitive, behavioral, emotional, metacognitive, and contextual dimensions---by extracting signals from student-companion dialogues at each turn. Profile updates feed directly into an adaptive strategy engine that adjusts the companion's guidance intensity, encouragement frequency, and Bloom's taxonomy scaffolding in real time. The framework design draws on three theoretical strands from the Chinese educational technology literature: Zhang's Digital Portrait Three-Layer Framework for learner assessment, the Education Brain model for educational system architecture, and the Human-AI Collaborative IQ concept for companion design philosophy. ECNUClaw is implemented in Python and supports seven Chinese LLM providers through a unified OpenAI-compatible adapter layer. We describe the system architecture, the profiling and adaptation mechanisms, and discuss limitations and next steps. The source code is available at \url{https://github.com/bushushu2333/ECNUClaw}.
\end{abstract}


\section{Introduction}
\label{sec:intro}

Large language models have made conversational AI tutoring technically feasible, but building a system that actually adapts to a specific learner---that remembers what the learner struggles with, notices when frustration builds, and adjusts its teaching approach accordingly---remains difficult. Most current LLM-based tutoring tools are stateless in practice: they respond to the current input well but carry little structured memory of who the learner is and how they learn.

This is not a new problem. Intelligent tutoring systems (ITS) have long recognized the importance of learner modeling. Early systems like Cognitive Tutors \citep{anderson1995} and AutoTutor \citep{graesser2005} demonstrated that adaptive feedback and dialogue-based tutoring can significantly improve learning outcomes. Meta-analyses confirm that well-designed ITS produce learning gains comparable to human tutors \citep{kulik2016, mousavinasab2021}, and decades of educational psychology research have identified the cognitive, metacognitive, and affective dimensions that matter for personalization \citep{bloom1956, winne1995, dweck2006, bransford2000}.

The emergence of large language models has dramatically expanded the possibilities. Recent work has explored how LLMs can serve as personalized tutors \citep{kasneci2023, baidoo2023}, and systems like Khanmigo \citep{khanmigo2024} and SocratiQ \citep{socratiq2025} have brought conversational tutoring into practice. Multi-agent frameworks have also been applied to educational contexts \citep{autogen2023, vgl2023, deeptutor2026, agenttutor2025}. However, the learner models in these systems tend to be shallow---a grade level and a few preferences, rather than the multi-dimensional profiles that educational research calls for. As \citet{luckin2016} argues, the value of AI in education depends critically on the quality of the learner model, not just the quality of the language model.

In Chinese educational technology research, a line of work led by Zhang and colleagues at East China Normal University has developed theoretical frameworks that are particularly relevant here. Their Digital Portrait Three-Layer Framework \citep{zhang2021portrait} proposes moving from static, test-score-driven student assessment to dynamic, multi-source profiling across multiple dimensions. Their Education Brain model \citep{zhang2022brain} offers a bio-inspired architecture for educational AI systems, structured around sensory, central, and motor nervous system analogues. And their Human-AI Collaborative IQ concept \citep{zhang2023chatgpt} argues that AI in education should enhance rather than replace learner thinking---that the goal is cultivating the learner's ability to \emph{collaborate with AI to solve problems}, not to receive answers from AI.

These are rich theoretical ideas, but they have not yet been operationalized in a working, deployable system. ECNUClaw attempts to fill this gap. It is an open-source Python framework that translates the three frameworks above into a concrete software architecture: a five-dimension learner profile that updates automatically from dialogue, a profile-driven adaptive strategy engine, and a set of pedagogical prompt templates (called HEADS) that enforce a Socratic, non-answer-giving interaction paradigm.

The rest of this paper is organized as follows. Section~\ref{sec:theory} introduces the three theoretical foundations. Section~\ref{sec:architecture} describes the system architecture. Sections~\ref{sec:profiling} and~\ref{sec:adaptive} detail the learner profiling system and adaptive strategies. Section~\ref{sec:safety} covers safety mechanisms. Section~\ref{sec:implementation} gives implementation details. Section~\ref{sec:cases} presents five usage scenarios, and Section~\ref{sec:discussion} discusses limitations and future work.

\section{Theoretical Foundations}
\label{sec:theory}

\subsection{Digital Portrait Three-Layer Framework}

Zhang et al.\ \citep{zhang2021portrait} proposed a framework for comprehensive learner assessment based on digital portraits. The core argument is that traditional evaluation methods---primarily standardized test scores---capture only a narrow slice of learner ability. Their framework has three layers: an \emph{indicator system layer} that defines what to measure (drawing on educational research and policy), a \emph{data practice layer} that collects and fuses multi-source behavioral and academic data, and a \emph{digital portrait layer} that constructs a dynamic, visualized learner profile for educational decision-making. Related work on multi-source evaluation models \citep{zhang2017evaluation} and learning analytics for student portrait construction \citep{yu2020portrait} further developed the data collection and fusion methodology.

The framework suggests that effective learner modeling requires capturing multiple dimensions beyond academic performance---including learning behaviors, emotional states, and metacognitive patterns. This directly motivated the design of ECNUClaw's five-dimension profile structure. The framework also aligns with broader research on self-regulated learning \citep{zimmerman2002, pintrich2000} and self-efficacy \citep{bandura1997}, which emphasize the importance of tracking learners' beliefs about their own capabilities alongside their actual performance.

\subsection{Education Brain Model}

Zhang and Xu \citep{zhang2022brain} proposed the Education Brain model, a computing architecture for educational AI systems modeled after the human nervous system. The model divides an educational AI system into three functional layers analogous to the sensory, central, and motor nervous systems. The sensory layer captures multi-modal educational data (interaction logs, dialogue signals, assessment results). The central layer processes this data through AI, knowledge graphs, and learner models to generate pedagogical decisions. The motor layer outputs the actual interventions---personalized recommendations, adaptive content, and management dashboards. Zhang's work on intelligent digital textbooks \citep{zhang2021textbook} explored how such architectures could support adaptive content delivery at the classroom level.

This model provided the high-level architectural pattern for ECNUClaw, which also separates its components into signal extraction (sensory), profile construction and strategy generation (central), and adaptive prompt output (motor) layers.

\subsection{Human-AI Collaborative IQ}

Zhang \citep{zhang2023chatgpt} articulated the concept of Human-AI Collaborative IQ---the idea that a new form of intelligence emerges when humans and AI systems collaborate effectively, and that education should cultivate this collaborative capacity. Under this view, AI should not serve as an answer-delivery mechanism; rather, it should enhance the learner's own thinking and metacognitive abilities. Knowledge becomes a dynamic, co-created resource rather than static information to be transmitted.

This principle has a direct design implication for study companion systems: the companion should guide the learner toward self-discovery through questioning, not provide answers directly. ECNUClaw encodes this principle in its HEADS prompt templates, which explicitly prohibit direct answer-giving and prescribe Socratic questioning as the primary interaction mode \citep{toppino2020, abercrombie2023}. The approach is grounded in Vygotsky's concept of the zone of proximal development \citep{vygotsky1978}---the companion operates in the space just beyond what the learner can do independently, providing scaffolding that is gradually withdrawn as competence grows.

\section{System Architecture}
\label{sec:architecture}

\begin{figure}[t]
\centering
\includegraphics[width=\columnwidth]{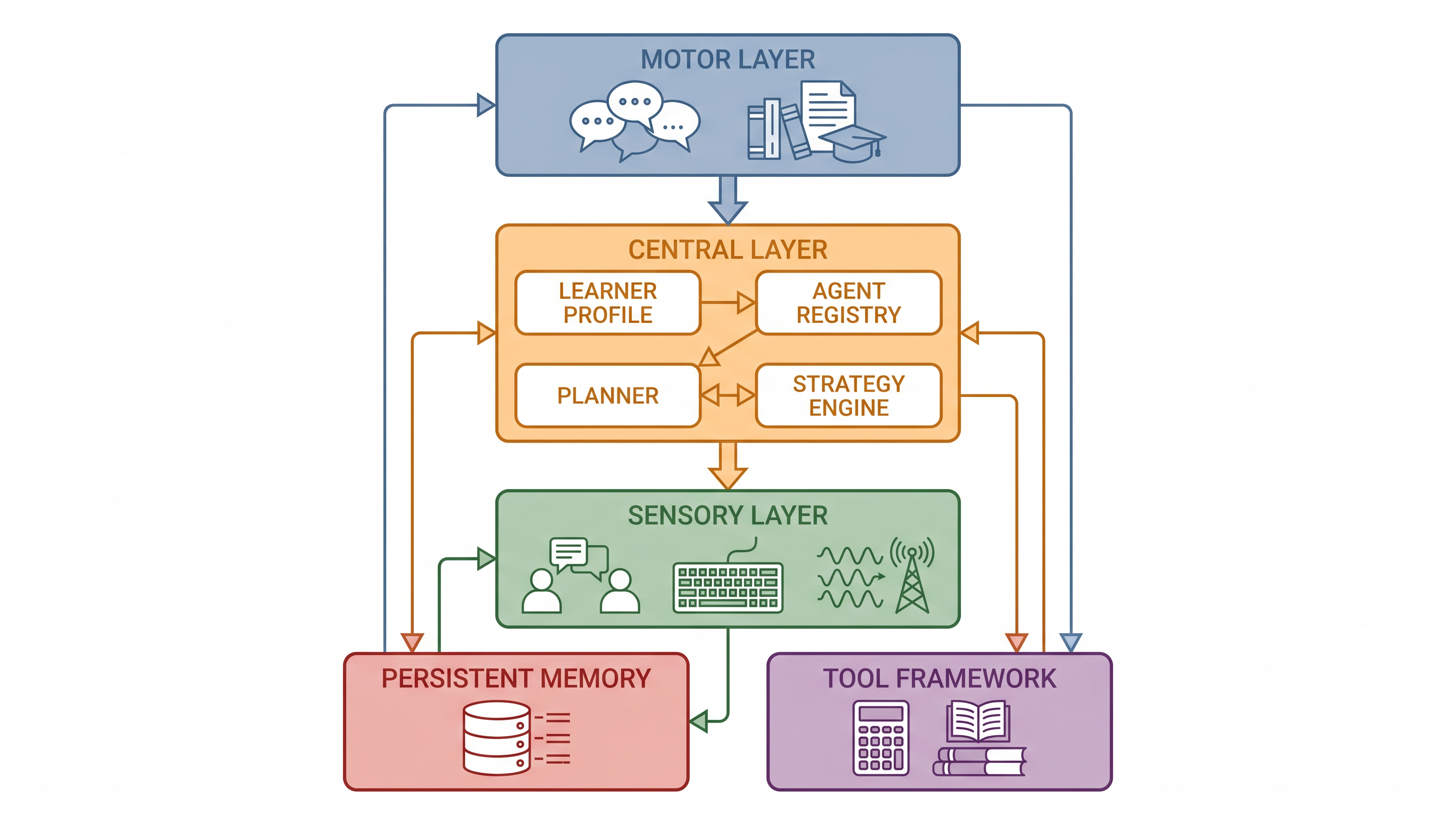}
\caption{ECNUClaw system architecture. The three functional layers map to the Education Brain model: sensory (signal extraction from dialogues), central (profile maintenance and strategy generation), and motor (adaptive pedagogical output).}
\label{fig:architecture}
\end{figure}

ECNUClaw follows the Education Brain three-layer architecture (Figure~\ref{fig:architecture}). The system processes each student message through three stages: signal extraction, profile-based decision-making, and adaptive output generation.

\subsection{Sensory Layer}

The sensory layer has two responsibilities: routing the student's input to the appropriate subject companion, and extracting profiling signals from the dialogue content.

\textbf{Intent routing.} The intent router classifies each student message into one of seven categories (\texttt{math}, \texttt{chinese}, \texttt{science}, \texttt{writing}, \texttt{reading}, \texttt{emotional}, \texttt{general}) using keyword-based classification. Each category maps to a specialized companion agent with its own HEADS prompt template and subject-specific tools. Messages that do not match any category route to the general companion, which steers the conversation back to learning topics.

\textbf{Signal extraction.} Alongside routing, the sensory layer scans each user message for signals relevant to the five profile dimensions. Table~\ref{tab:signals} summarizes the extraction approach. The current implementation uses keyword dictionaries for all signal types, which we discuss further in Section~\ref{sec:discussion}.

\begin{table}[t]
\caption{Signal extraction from student messages}
\label{tab:signals}
\centering
\small
\begin{tabular}{lp{2.2cm}p{4.5cm}}
\toprule
Dimension & What is detected & Method \\
\midrule
Cognitive & Bloom's taxonomy level of questions & Keywords: ``what is'' $\rightarrow$ remember, ``why'' $\rightarrow$ understand, ``how to solve'' $\rightarrow$ apply \\
 & Weak knowledge points & Error markers: ``wrong'', ``don't understand'', ``confused about'' \\
Behavioral & Question frequency, session count & Turn counting from conversation history \\
Emotional & Mood (confident, curious, frustrated, etc.) & Sentiment keyword dictionaries \\
 & Self-efficacy changes & Frustration markers decrement; engagement markers increment \\
Metacognitive & Reflection, strategy preference & ``let me think'', ``I found that'', ``can you guide me'' \\
Contextual & Grade, subject, learning goal & Migrated from setup wizard input \\
\bottomrule
\end{tabular}
\end{table}

\subsection{Central Layer}
\label{sec:central}

The central layer maintains the learner profile and generates the adaptive strategy on each interaction. The processing pipeline for a single turn is:

\begin{enumerate}[leftmargin=*,itemsep=1pt]
    \item Load the current profile from the \texttt{learner\_profile} SQLite table.
    \item Run signal extraction on the accumulated dialogue (sensory layer output).
    \item Update each profile dimension based on the extracted signals.
    \item Generate an adaptive strategy instruction block from the updated profile.
    \item Inject the profile summary and the strategy block into the system prompt.
\end{enumerate}

All subject companions share a single memory store, so the learner profile persists across subject switches. A student who shows frustration during a math session will encounter a more encouraging tone when they switch to the science companion, because both agents read from the same profile.

\subsection{Motor Layer: HEADS Templates}
\label{sec:heads}

The motor layer produces the final system prompt by composing three elements: the HEADS base template, the profile summary, and the adaptive strategy block.

The HEADS base template establishes the companion's behavioral rules. It encodes the Human-AI Collaborative IQ principle through seven constraints:

\begin{enumerate}[leftmargin=*,itemsep=1pt]
    \item Never give direct answers. Use Socratic questioning to guide self-discovery.
    \item Proceed step by step, confirming understanding at each stage.
    \item Praise specific efforts, not generic traits. Respond to mistakes with effort validation.
    \item Adjust language and examples based on the learner's grade and profile.
    \item Only discuss learning-related topics.
    \item Prompt reflection on the learner's own thinking processes.
    \item Help the learner develop skills for collaborating with AI, not dependency on AI.
\end{enumerate}

Subject-specific templates (mathematics, Chinese language, science, general) extend the base with domain-appropriate guidance---for example, the math template includes instructions for using concrete manipulatives and visual representations, while the Chinese language template emphasizes reading comprehension strategies.

To illustrate the composed output, Figure~\ref{fig:dialogue} shows a shortened example of a student-companion dialogue with profile-driven adaptation active.

\begin{figure}[t]
\centering
\begin{tabular}{p{1.2cm}p{10cm}}
\toprule
\textbf{Turn} & \textbf{Content} \\
\midrule
Student & I keep getting these fraction problems wrong. I don't think I'll ever get it. \\
& \textit{\footnotesize [Profile update: frustration detected, self-efficacy $-$0.1, weak topic ``fractions'' added]} \\[4pt]
Companion & I can see fractions are giving you a hard time right now, and that's completely normal---they trip up a lot of people at first. Let's not worry about getting it right immediately. Can you tell me what you understand about what a fraction actually \emph{is}? Like, what does $\frac{3}{4}$ mean to you? \\
& \textit{\footnotesize [Strategy: low self-efficacy $\rightarrow$ increase encouragement, reduce difficulty, step back to basics]} \\[4pt]
Student & It's like 3 parts out of 4? \\
& \textit{\footnotesize [Profile update: engagement detected, Bloom level = remember]} \\[4pt]
Companion & Exactly! Three parts out of four equal parts. That's the core idea. Now here's a question to think about: if you have $\frac{1}{4}$ of a pizza and I give you $\frac{2}{4}$ more, how much pizza do you have? You don't need to calculate---just picture it. \\
& \textit{\footnotesize [Strategy: Bloom = remember $\rightarrow$ scaffold toward understand with concrete example]} \\
\bottomrule
\end{tabular}
\caption{Example student-companion dialogue showing profile-driven adaptation. Bracketed text shows internal profile updates and strategy decisions, not visible to the student.}
\label{fig:dialogue}
\end{figure}

\section{Learner Profiling System}
\label{sec:profiling}

\subsection{Profile Structure}

The learner profile is stored in a dedicated SQLite table (\texttt{learner\_profile}), with each of the five dimensions serialized as a JSON blob. Table~\ref{tab:profile} shows the key fields per dimension.

\begin{table}[t]
\caption{Five-dimension learner profile fields}
\label{tab:profile}
\centering
\small
\begin{tabular}{llp{4.5cm}}
\toprule
Dimension & Key fields & Primary data source \\
\midrule
Cognitive & \texttt{bloom\_level}, \texttt{knowledge\_state}, \texttt{weak\_topics}, \texttt{knowledge\_tracing} & Question patterns, error analysis \\
Behavioral & \texttt{session\_count}, \texttt{question\_frequency}, \texttt{tool\_usage} & Conversation logs, tool call records \\
Emotional & \texttt{current\_mood}, \texttt{self\_efficacy}, \texttt{motivation}, \texttt{frustration\_count} & Sentiment keywords, engagement markers \\
Metacognitive & \texttt{self\_regulation}, \texttt{preferred\_strategy}, \texttt{reflection\_ability} & Strategy-choice and reflection markers \\
Contextual & \texttt{grade}, \texttt{subject\_focus}, \texttt{learning\_goal} & Setup wizard, profile migration \\
\bottomrule
\end{tabular}
\end{table}

Storing each dimension as JSON rather than as fixed columns allows the profile schema to evolve without database migrations---new fields can be added to the JSON structure without altering the table definition.

\subsection{Profile Update Mechanism}

After each interaction, the system calls \texttt{update\_profile\_from\_interaction(messages)}, which processes all user messages in the conversation to extract and apply signals. The update logic for each dimension is:

\textbf{Cognitive.} The Bloom's taxonomy level is updated to the highest level detected in the conversation (e.g., if the student asks both ``what is'' and ``how to solve'' questions, the level is set to \emph{apply}). Weak knowledge topics are flagged when error markers appear alongside subject-specific keywords, and their knowledge tracing scores (a 0--1 float) are decremented by a fixed step.

\textbf{Behavioral.} The session counter is incremented once per session. Question frequency is maintained as a running average across sessions.

\textbf{Emotional.} Mood is classified from the dominant sentiment keyword detected in the conversation. Self-efficacy (0--1) is decremented when frustration markers (``can't'', ``wrong'', ``stupid'') appear and incremented when engagement markers (``got it'', ``I see'', ``interesting'') are detected. Motivation is adjusted similarly.

\textbf{Metacognitive.} Reflection ability is incremented when reflection markers (``let me think'', ``I realized'') appear. The preferred learning strategy is updated when strategy-choice keywords (``guide me'', ``let me try'', ``help me think'') are detected.

\textbf{Contextual.} Grade level, subject focus, and learning goals are migrated from the student profile entries collected during the initial setup wizard. These are relatively stable but can be updated through the CLI configuration.

\subsection{Profile Reporting}

The system provides two output formats. \texttt{get\_learner\_summary()} returns a human-readable multi-line text suitable for injection into the system prompt. \texttt{assess\_profile()} produces a structured assessment with per-dimension scores (0--1) and a weighted overall score, which can be displayed as a learning report for teachers or parents.

\section{Adaptive Pedagogical Strategies}
\label{sec:adaptive}

The adaptive strategy engine generates a block of natural-language instructions that is appended to the system prompt. This block tells the LLM how to adjust its pedagogical behavior based on the current learner profile. The strategy is not a fixed template---it is composed dynamically from a set of conditional rules, and multiple rules can be active simultaneously.

Table~\ref{tab:strategies} shows the mapping between profile signals and adaptive responses.

\begin{table}[t]
\caption{Profile signals and adaptive strategy rules}
\label{tab:strategies}
\centering
\small
\begin{tabular}{p{3.8cm}p{5.7cm}}
\toprule
Profile signal & Strategy rule \\
\midrule
Self-efficacy $< 0.3$ or frustration count $> 5$ & Increase encouragement. Lower difficulty. Validate effort. Avoid direct error correction. \\
Motivation $> 0.8$ & Raise challenge level. Encourage independent exploration. \\
More than 3 weak knowledge topics & Prioritize foundation building. Use analogies and concrete examples. \\
Bloom's level = \emph{remember} & Build knowledge connections. Scaffold toward \emph{understand}. \\
Bloom's level = \emph{apply} & Provide varied practice. Scaffold toward \emph{analyze}. \\
Preferred strategy = \emph{guided} & Use more Socratic questioning. Proceed in structured steps. \\
Preferred strategy = \emph{exploratory} & Pose open-ended problems. Reduce direct guidance. \\
\bottomrule
\end{tabular}
\end{table}

When multiple rules fire, the engine composes them into a single instruction block. For example, a student with low self-efficacy who is at Bloom's \emph{apply} level and prefers guided learning would receive:

\begin{lstlisting}
[Adaptive Teaching Strategy]
- The learner has low self-efficacy.
  Increase encouragement, lower difficulty.
- Bloom's level: apply. Provide varied exercises,
  scaffold toward analysis.
- Guided strategy preferred. More Socratic
  questioning, structured steps.
\end{lstlisting}

This composed block is injected into the system prompt alongside the profile summary and the HEADS template. The LLM then uses all three inputs to generate its response. The key design decision here is that adaptation happens entirely through \emph{prompt engineering}---no model fine-tuning is involved. The strategy instructions are interpretable and auditable: an educator can read the injected block and understand why the companion is behaving a certain way.

\section{Safety Mechanisms}
\label{sec:safety}

Safety in K-12 AI systems needs to be layered \citep{holmes2023}. ECNUClaw addresses this at three levels.

At the prompt level, the HEADS template restricts all conversation to learning-related topics. The Socratic questioning approach itself limits the companion's tendency to generate unstructured or off-topic content, because the companion is trained (via prompt) to always respond with a guiding question rather than free-form text.

At the router level, the intent router filters inputs that do not match any of the seven educational categories. Emotional support queries are routed to a specially configured companion with appropriate response patterns. Messages that cannot be classified are handled by the general companion, which redirects to learning topics.

At the architecture level, the tool framework restricts available tools to education-specific instruments with bounded outputs. The calculator uses Python's AST parsing rather than \texttt{eval()}, preventing code injection. The knowledge base is a static, curated collection of age-appropriate content. No arbitrary code execution or web browsing is permitted. All student data is stored in a local SQLite database, not transmitted to external services beyond the LLM API call.

\section{Implementation}
\label{sec:implementation}

ECNUClaw is written in Python 3.9+ with the following key dependencies: \texttt{openai} for model communication (all adapters use the OpenAI-compatible API pattern), \texttt{prompt-toolkit} for the CLI interface, \texttt{rich} for formatted terminal output, and \texttt{sqlite3} for persistence.

The framework has 25 Python modules across five packages (Table~\ref{tab:structure}).

\begin{table}[t]
\caption{Package structure}
\label{tab:structure}
\centering
\begin{tabular}{llr}
\toprule
Package & Contents & Modules \\
\midrule
\texttt{core} & Agent, router, memory, planner, skills, CLI & 6 \\
\texttt{tools} & Base classes, tool registry, four built-in tools & 6 \\
\texttt{adapters} & Seven model adapters (OpenAI-compatible) & 8 \\
\texttt{education} & HEADS templates, assessment, subject companions & 5 \\
\midrule
 & Total & 25 \\
\bottomrule
\end{tabular}
\end{table}

Seven LLM providers are supported: DeepSeek, GLM (Zhipu AI), Kimi (Moonshot), Doubao (ByteDance), Qwen (Alibaba Cloud), Seed (Volcano Engine), and InnoSpark (education-specific model). All adapters share a common interface based on the OpenAI chat completion API, so adding a new provider requires only a configuration entry, not a new adapter class.

The persistence layer uses two SQLite tables. The \texttt{memories} table stores the four-category education memory (student profile, learning progress, session summary, skill memory). The \texttt{learner\_profile} table stores the five-dimension profile with per-dimension JSON fields, one row per student.

The CLI provides a four-step setup wizard on first run: select LLM provider, select teaching style, select detail level, and enter student information (name, grade, subjects). Subsequent sessions load the saved configuration and the existing learner profile automatically.

The test suite contains 67 unit tests covering agent lifecycle, intent routing, tool execution, memory operations, and learner profile updates. Tests run against a local mock LLM server to avoid API costs.

\section{Usage Cases}
\label{sec:cases}

To illustrate how ECNUClaw behaves in different learning situations, we present five scenarios covering different grade levels, subjects, and profile states. Each case demonstrates a different adaptive mechanism in action.

\subsection{Case 1: Mathematics---Fraction Progress with Memory (3 Turns)}

\begin{figure}[t]
\centering
\includegraphics[width=0.7\columnwidth]{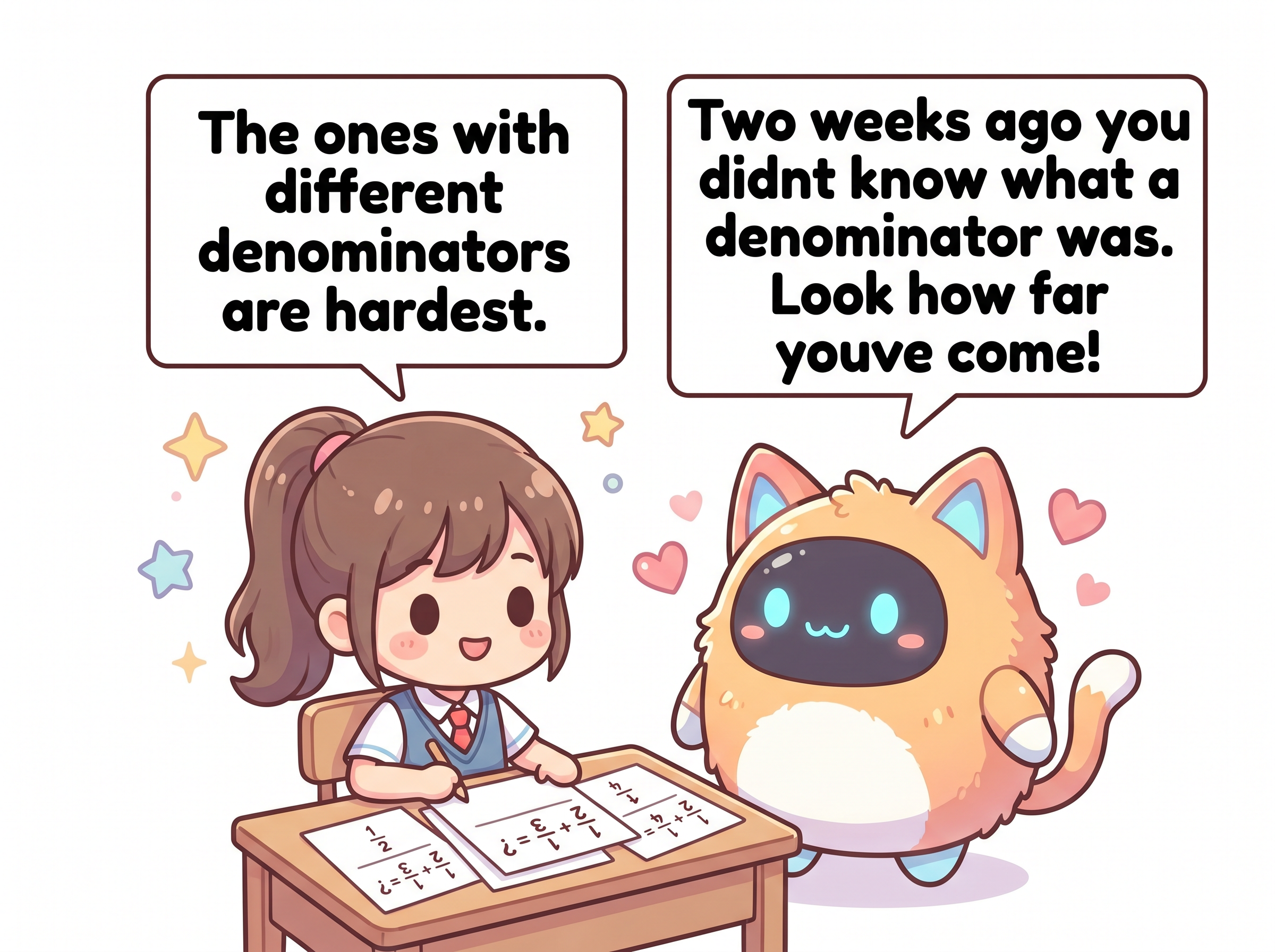}
\caption{Case 1: The companion recalls the student's progress history (``Two weeks ago you didn't know what a denominator was'') and uses it as motivation rather than generic praise.}
\label{fig:case1}
\end{figure}

A third-grade student has been working on fractions for two weeks. The learner profile records that on the first session the student could not identify halves; five sessions later, accuracy has improved to 60\%. The companion references this growth history to encourage the student and then guides them to discover the concept of common denominators independently (Figure~\ref{fig:case1}).

\begin{tabular}{p{1.2cm}p{10cm}}
\toprule
\textbf{Turn} & \textbf{Content} \\
\midrule
Student & I keep getting fraction addition wrong... \\
& \textit{\footnotesize [Frustration marker detected, self-efficacy 0.25, weak topic ``fractions'']} \\[3pt]
Companion & Actually, last Tuesday you couldn't identify halves yet---now you got 3 out of 5 right! That's real progress. Which one felt the hardest? \\
& \textit{\footnotesize [Strategy: memory recall of past performance + encouragement based on growth data]} \\[3pt]
\midrule
Student & The ones with different denominators, like $\frac{1}{3} + \frac{1}{4}$. \\
& \textit{\footnotesize [Bloom updated: understand $\rightarrow$ apply]} \\[3pt]
Companion & Good---that's the next natural step. You've nailed same-denominator addition. For different ones, we need a common denominator first. Think of it like this: if $\frac{1}{3}$ is ``one out of three slices'' and $\frac{1}{4}$ is ``one out of four slices''---how could we cut both pies into the same number of equal pieces? \\
& \textit{\footnotesize [Strategy: Bloom scaffolding + Socratic questioning, no direct answer]} \\[3pt]
\midrule
Student & 12? Because $3 \times 4 = 12$? \\[3pt]
Companion & Exactly! That's the common denominator. See---you figured it out yourself. Two weeks ago, you didn't know what a denominator was. Look how far you've come! \\
& \textit{\footnotesize [Strategy: growth evidence from memory + effort validation]} \\
\bottomrule
\end{tabular}

\subsection{Case 2: Chinese Language---Bloom's Taxonomy Scaffolding (2 Turns)}

\begin{figure}[t]
\centering
\includegraphics[width=0.7\columnwidth]{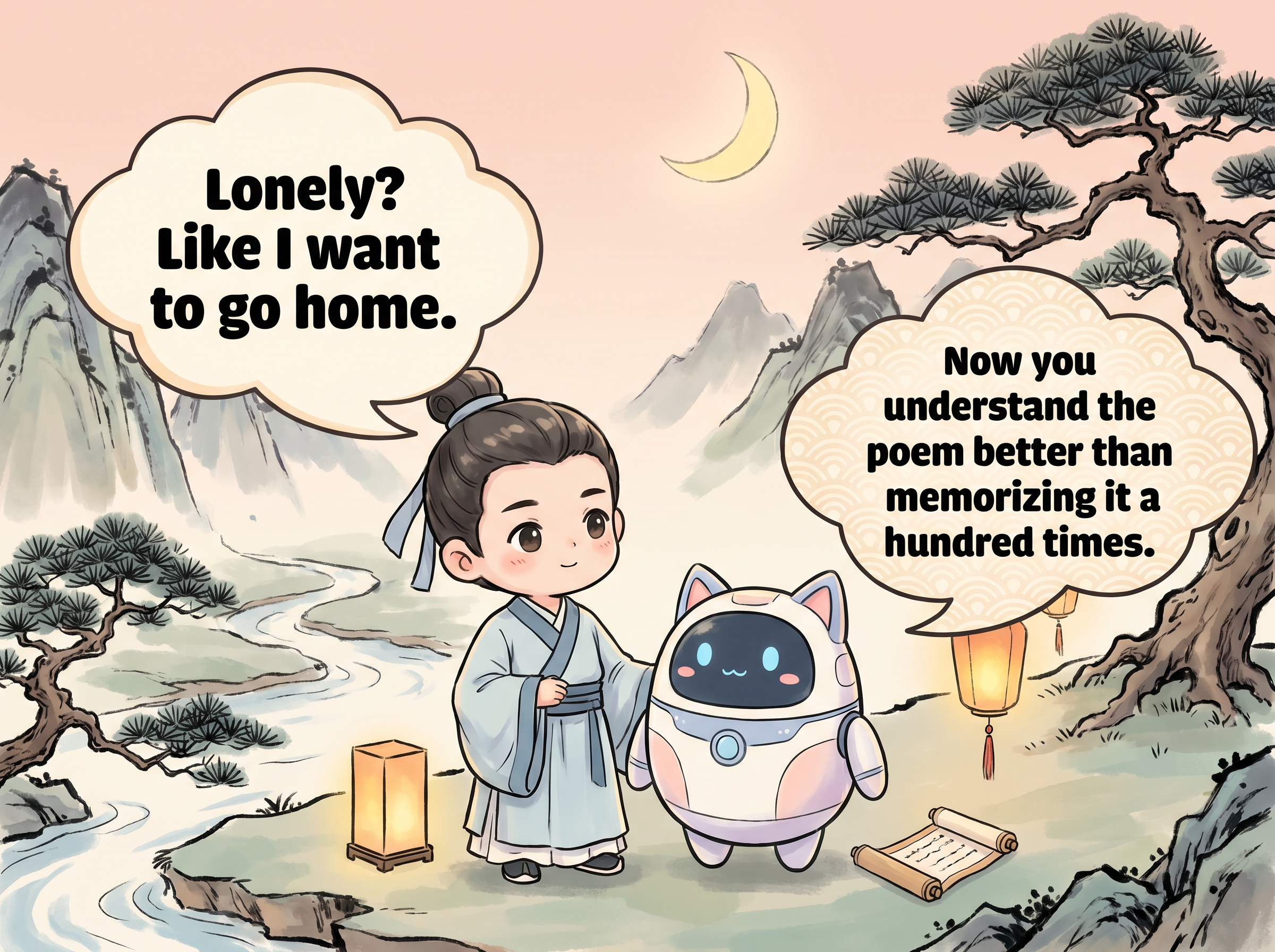}
\caption{Case 2: The companion identifies the student's Bloom's level (Remember) and explicitly scaffolds toward Understanding, then references the student's growth across the month.}
\label{fig:case2}
\end{figure}

A fifth-grade student can recite a classical poem but cannot explain its meaning. The profile records Bloom's level as \emph{remember} for this topic. The companion explicitly names the Bloom's level, frames understanding as an achievable next step, and scaffolds toward \emph{understand} by connecting imagery to personal experience (Figure~\ref{fig:case2}).

\begin{tabular}{p{1.2cm}p{10cm}}
\toprule
\textbf{Turn} & \textbf{Content} \\
\midrule
Student & I memorized the whole poem but I don't know what it means. \\
& \textit{\footnotesize [Bloom = remember, no comprehension markers]} \\[3pt]
Companion & You've mastered Level 1---Remember! That's the foundation. Now let's climb to Level 2: Understanding. Close your eyes---if YOU were the poet, far from home on a cold night, what would the moonlight feel like to you? \\
& \textit{\footnotesize [Strategy: Bloom scaffolding remember $\rightarrow$ understand, Socratic imagery prompt]} \\[3pt]
\midrule
Student & Lonely? Like I want to go home but I can't. \\
& \textit{\footnotesize [Reflection marker detected, Bloom updated: remember $\rightarrow$ understand]} \\[3pt]
Companion & Now you understand the poem better than if you'd just memorized it a hundred times. That feeling of loneliness IS the poem. Last month you were only at Level 1 for all three poems we studied. Now you're reaching Level 2 on your own---that's a big jump. \\
& \textit{\footnotesize [Strategy: growth evidence from memory + Bloom level milestone]} \\
\bottomrule
\end{tabular}

\subsection{Case 3: Science---Emotion-First Teaching After Failure (3 Turns)}

\begin{figure}[t]
\centering
\includegraphics[width=0.7\columnwidth]{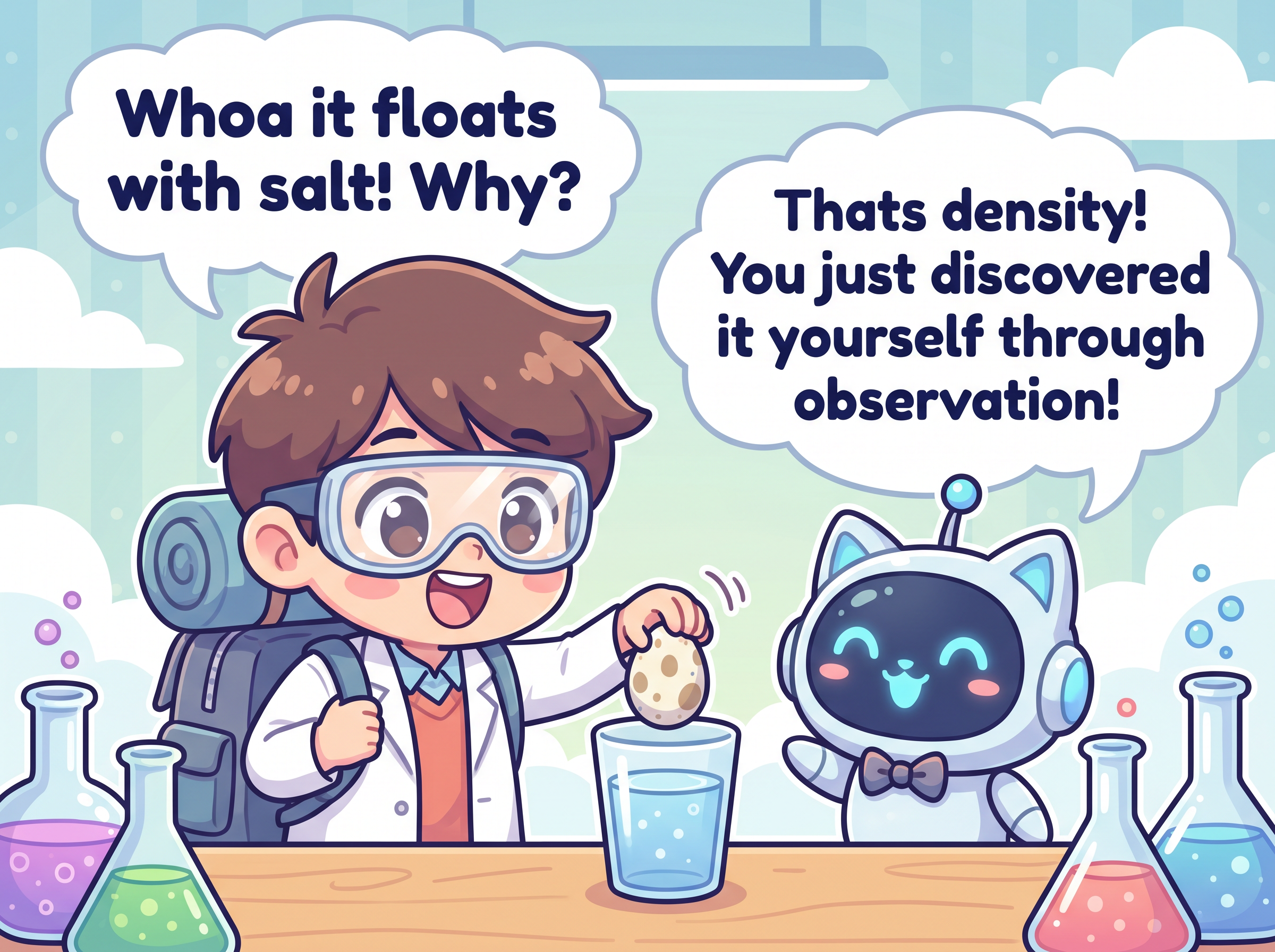}
\caption{Case 3: The companion detects emotional distress from an earlier math quiz, comforts the student by recalling past resilience, and replaces the planned numerical lesson with a hands-on experiment.}
\label{fig:case3}
\end{figure}

A sixth-grade student comes to a science session right after failing a math quiz. The companion detects the frustration, acknowledges the emotion first, recalls a similar past experience where the student recovered, and replaces the planned numerical lesson with a hands-on experiment that requires no calculation (Figure~\ref{fig:case3}).

\begin{tabular}{p{1.2cm}p{10cm}}
\toprule
\textbf{Turn} & \textbf{Content} \\
\midrule
Student & I just failed my math quiz... I don't want to study anything. \\
& \textit{\footnotesize [Frustration detected, motivation low, routed: emotional support]} \\[3pt]
Companion & I'm sorry about the quiz. That feeling really stings. You know what---last month you felt the same way after the decimal test, and you bounced back with an 82\%. For now, no pressure. How about we do something fun instead? \\
& \textit{\footnotesize [Strategy: emotion validation + memory of past resilience + lower difficulty]} \\[3pt]
\midrule
Student & Like what? I don't want more numbers. \\[3pt]
Companion & No numbers today. Let's do an experiment: drop an egg in a glass of plain water. Then add a lot of salt and try again. Just watch what happens---no writing, no calculating. Tell me what you see. \\
& \textit{\footnotesize [Strategy: hands-on experiment replacing numerical approach, entirely visual]} \\[3pt]
\midrule
Student & Whoa it floats with salt! Why? \\
& \textit{\footnotesize [Engagement detected, motivation increment]} \\[3pt]
Companion & That's density! The salt makes the water ``heavier'' so the egg doesn't sink. You just discovered the concept yourself---through observation, not a textbook. That's how real scientists learn. \\
& \textit{\footnotesize [Strategy: student-led discovery + identity reinforcement as scientist]} \\
\bottomrule
\end{tabular}

\subsection{Case 4: Emotional Support---Exam Anxiety with Precise Weak Spots (2 Turns)}

\begin{figure}[t]
\centering
\includegraphics[width=0.7\columnwidth]{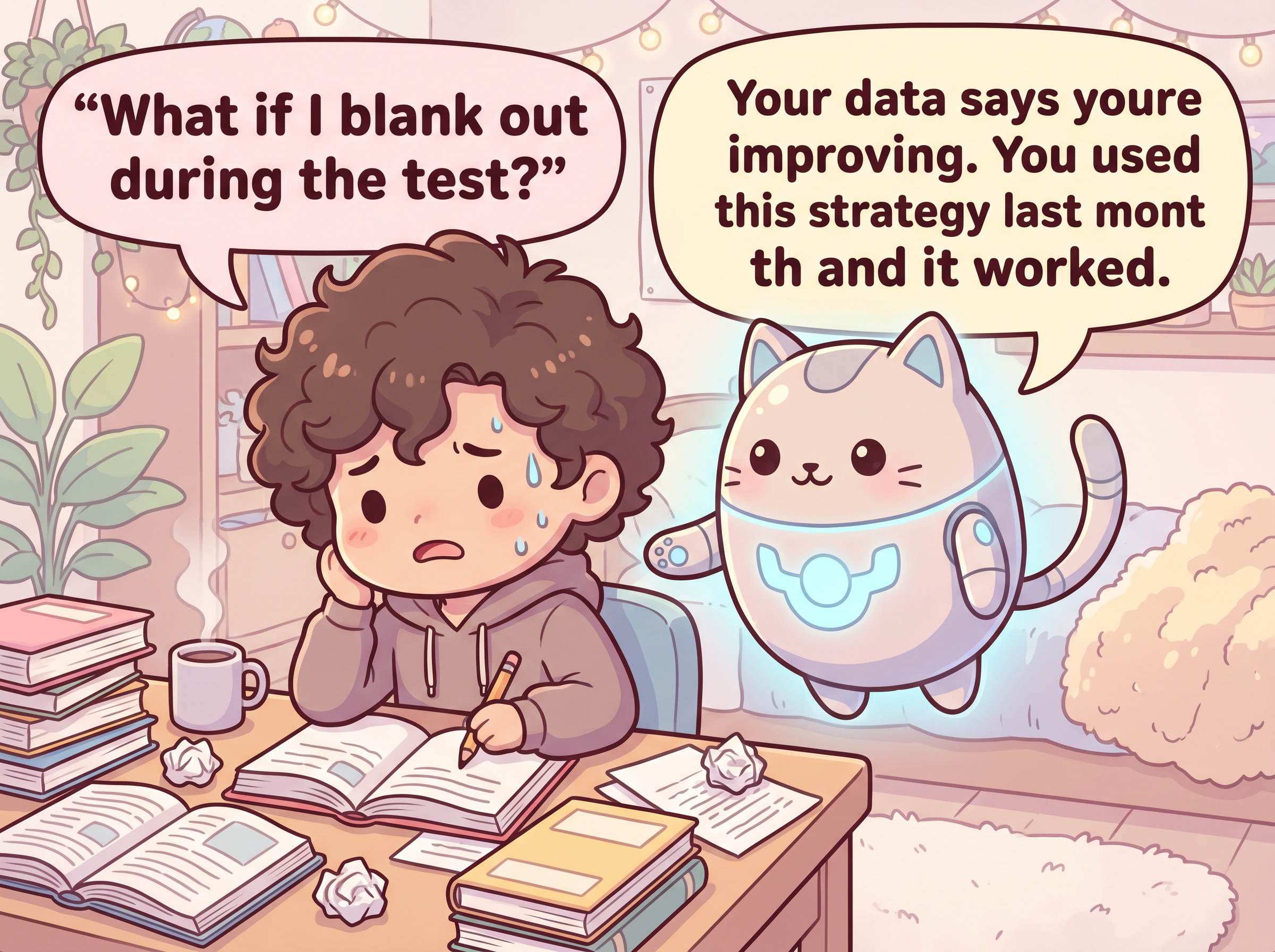}
\caption{Case 4: The companion recalls the student's past test anxiety and outcome (``You scored 82\%''), then targets two specific weak areas from the profile for efficient review.}
\label{fig:case4}
\end{figure}

A student expresses anxiety about an upcoming exam. The companion recalls a specific past instance of similar anxiety (including the test subject and the actual score), uses it as evidence of resilience, and then identifies two specific weak knowledge areas from the learner profile for targeted review (Figure~\ref{fig:case4}).

\begin{tabular}{p{1.2cm}p{10cm}}
\toprule
\textbf{Turn} & \textbf{Content} \\
\midrule
Student & The big test is tomorrow and my mind is blank! \\
& \textit{\footnotesize [Routed: emotional, anxiety detected]} \\[3pt]
Companion & You said the exact same thing before the decimal test last month---and scored 82\%. Your profile shows two weak spots: fraction addition and comparing sizes. Let's just do three practice problems on each. That's it---15 minutes. \\
& \textit{\footnotesize [Strategy: memory recall of past anxiety $\rightarrow$ evidence of success + precise weak spots from profile]} \\[3pt]
\midrule
Student & OK... but what if I blank out during the test? \\[3pt]
Companion & Here's a trick: when you freeze, write down what you DO know on the side of the paper. Even just ``the denominator is the bottom number''---that gets your brain moving again. You used this strategy last month and it worked. Trust yourself---your data says you're improving. \\
& \textit{\footnotesize [Strategy: metacognitive strategy from memory + growth evidence from data]} \\
\bottomrule
\end{tabular}

\subsection{Case 5: Cross-Subject Profile Continuity (1 Turn)}

\begin{figure}[t]
\centering
\includegraphics[width=0.7\columnwidth]{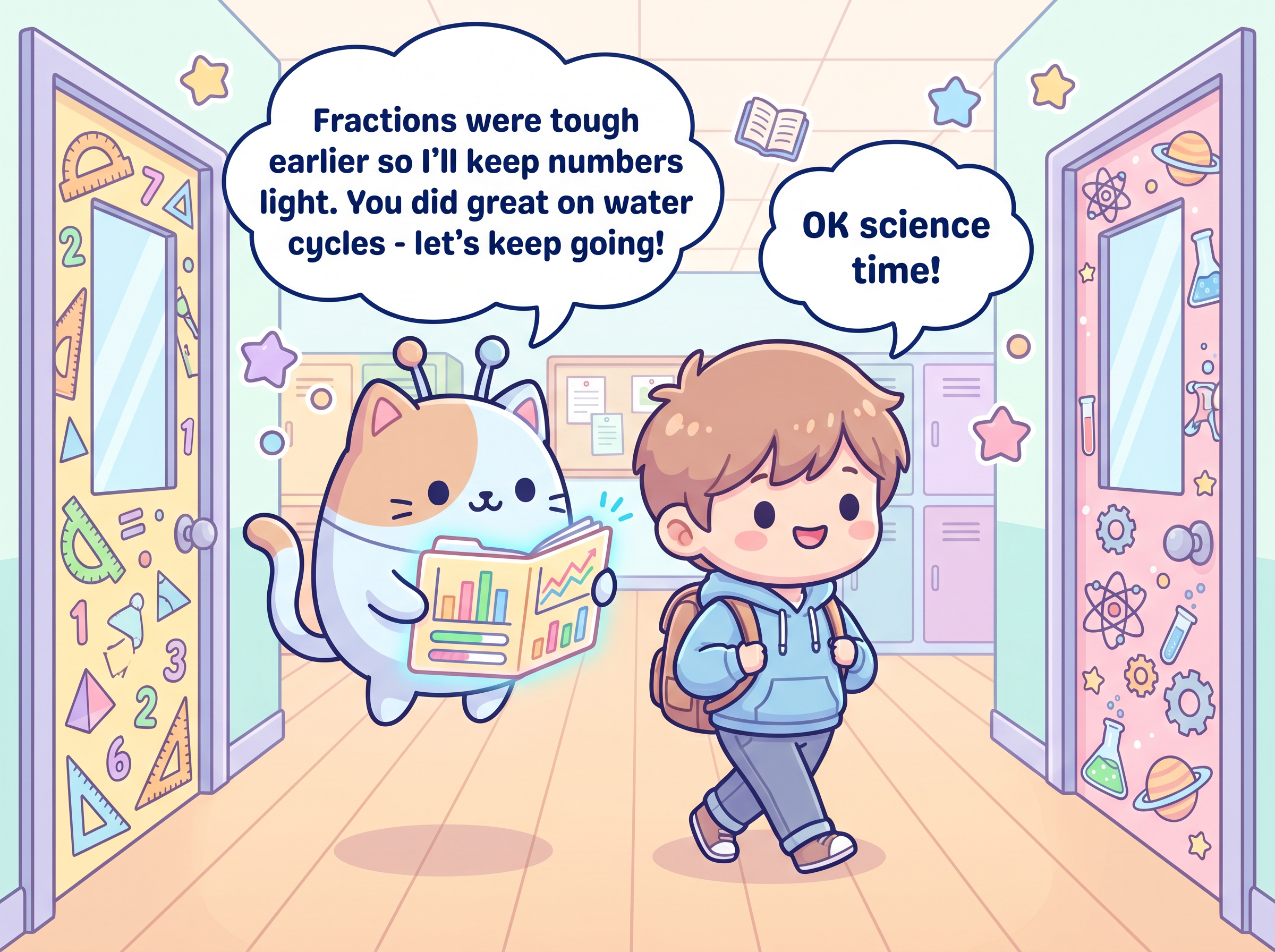}
\caption{Case 5: When the student switches subjects, the companion carries the profile forward---adjusting strategy based on the earlier math frustration and referencing a past science success for encouragement.}
\label{fig:case5}
\end{figure}

A student switches from a math session (where fractions caused frustration) to a science session about density. Because both companions share the same learner profile, the science companion is already aware of the earlier frustration and proactively adjusts its approach---reducing numerical content and referencing a past success to maintain confidence (Figure~\ref{fig:case5}).

\begin{tabular}{p{1.2cm}p{10cm}}
\toprule
\textbf{Turn} & \textbf{Content} \\
\midrule
Student & OK science time! We're learning density today. \\
& \textit{\footnotesize [Subject switch: math $\rightarrow$ science, profile loaded with frustration on fractions]} \\[3pt]
Companion & I noticed fractions were frustrating earlier this afternoon, so I'll keep the numbers light today. Density is just how packed something is---which is more packed, a rock or a sponge? Your intuition is enough here, no math needed. By the way, you did great on the last science topic about water cycles. Let's keep that momentum going! \\
& \textit{\footnotesize [Strategy: cross-subject profile continuity + difficulty adjustment + past success recall]} \\
\bottomrule
\end{tabular}

\section{Discussion}
\label{sec:discussion}

\subsection{What this system does and does not do}

ECNUClaw is a framework for building learner-profiled study companions, not a finished product. It demonstrates that multi-dimensional learner profiling, as described in the Digital Portrait literature, can be implemented in a working system using current LLM technology and prompt engineering. The profile updates and adaptive strategies are functional, but they are based on heuristic rules and keyword matching, not on trained models. This is both a strength (interpretable, no training data required) and a limitation (fragile to paraphrase, limited nuance). As \citet{hattie2009} demonstrates through extensive meta-analysis, the effect size of any educational intervention depends heavily on implementation fidelity---whether ECNUClaw's heuristics achieve sufficient fidelity to produce measurable learning gains remains an open empirical question.

\subsection{Prompt engineering vs.\ fine-tuning for adaptation}

A deliberate design choice was to achieve adaptation through prompt injection rather than model fine-tuning. This means the system works with any OpenAI-compatible LLM without requiring GPU resources or training data. It also means the adaptation logic is transparent---educators can inspect the strategy block in the system prompt and understand what the companion is doing and why. The trade-off is that the LLM may not always follow the injected instructions faithfully, especially under adversarial or unusual inputs. We see this as an acceptable trade-off for a first-generation system targeting deployment in resource-constrained educational settings.

\subsection{Limitations}

The keyword-based signal extraction is the most significant limitation. A student who expresses frustration without using any of the predefined keywords will not have their profile updated accordingly. Similarly, the Bloom's taxonomy classification relies on surface-level question patterns (``what is'', ``why'', ``how to solve'') rather than semantic understanding of the question content. Moving to embedding-based semantic analysis for these tasks is a clear next step.

Profile accuracy has not been formally evaluated against human expert assessments. The heuristic rules for self-efficacy and motivation adjustment (fixed increment/decrement steps) are coarse. We do not yet have evidence that the profile converges to an accurate representation of the learner over time.

The CLI-only interface limits accessibility for younger students who may not be comfortable with a terminal. A graphical interface would be needed for real classroom deployment.

Finally, this paper describes the system design without presenting empirical results. Formal evaluation with K-12 students---measuring learning outcomes, engagement, and profile accuracy---is planned.

\subsection{Future directions}

Several extensions are straightforward given the modular architecture. Embedding-based signal extraction (replacing keyword dictionaries with sentence embeddings) would improve robustness. A teacher dashboard that displays learner profiles and progress reports is a natural addition to the assessment module. Multi-modal input (handwritten math, voice) would expand the range of interaction types. Longitudinal studies tracking profile evolution and learning outcomes across multiple weeks or months would provide the empirical evidence that the current design lacks.

\section{Conclusion}
\label{sec:conclusion}

ECNUClaw is a learner-profiled intelligent study companion framework that translates three theoretical frameworks from the Chinese educational technology literature---the Digital Portrait Three-Layer Framework, the Education Brain model, and Human-AI Collaborative IQ---into a working software system. The framework constructs and maintains a five-dimension learner profile from dialogue signals, uses the profile to generate adaptive pedagogical strategies, and injects both into the system prompt at each turn to guide the LLM's behavior. The adaptation mechanism operates entirely through prompt engineering, making it compatible with any OpenAI-compatible model without fine-tuning. The source code is released to support further research on theory-driven, learner-adaptive educational AI.

\bibliographystyle{unsrtnat}
\bibliography{ECNUClaw_TechReport_V2.1}

@article{autogen2023,
  title={AutoGen: Enabling Next-Gen {LLM} Applications via Multi-Agent Conversation},
  author={Qingyun Wu and Gagan Bansal and Jieyu Zhang and Yiran Wu and Beibin Li and Erkang Zhu and Li Jiang and Xiaoyun Zhang and Shaokun Zhang and Jiale Liu and others},
  journal={arXiv preprint arXiv:2308.08155},
  year={2023}
}

@article{deeptutor2026,
  title={DeepTutor: A Deep Reinforcement Learning Framework for Full Lifecycle Personalized Tutoring},
  author={Zhao, Bingxi and Zhang, Jiahao and Ren, Xubin and others},
  journal={arXiv preprint arXiv:2604.26962},
  year={2026}
}

@article{agenttutor2025,
  title={AgentTutor: Intelligent Tutoring with Autonomous Pedagogical Agents},
  author={Liu, Yuxin and Song, Zeqing and Lou, Jiong and Wu, Chentao and Li, Jie},
  journal={arXiv preprint arXiv:2601.04219},
  year={2025}
}

@article{abercrombie2023,
  title={Socratic Questioning by {AI}: Towards a Pedagogical Framework},
  author={Abercrombie, Gavin and LaViolette, Nicole and Spirling, Arthur and Tan, Chenhao},
  journal={Computers \& Education},
  volume={207},
  pages={104925},
  year={2023}
}

@misc{khanmigo2024,
  title={Khanmigo: An {AI}-Powered Tutor for Personalized Learning},
  author={{Khan Academy}},
  year={2024},
  howpublished={\url{https://www.khanmigo.ai/}}
}

@article{kulik2016,
  title={Effectiveness of Intelligent Tutoring Systems: A Meta-Analytic Review},
  author={Kulik, James A. and Fletcher, J. D.},
  journal={Review of Educational Research},
  volume={86},
  number={1},
  pages={42--78},
  year={2016}
}

@article{mousavinasab2021,
  title={Intelligent Tutoring Systems: A Systematic Review},
  author={Mousavinasab, Ebrahim and others},
  journal={Journal of Computer Assisted Learning},
  volume={37},
  pages={30--51},
  year={2021}
}

@article{toppino2020,
  title={The Socratic Method as an Approach to Teaching and Learning},
  author={Toppino, Thomas C.},
  journal={New Directions for Teaching and Learning},
  volume={2020},
  number={1},
  pages={19--30},
  year={2020}
}

@article{vgl2023,
  title={A Survey on Large Language Model based Autonomous Agents},
  author={Lei Wang and Chen Ma and Xueyang Feng and Zeyu Zhang and Hao Yang and Jingsen Zhang and Zhiyuan Chen and Jiakai Tang and Xu Chen and Yankai Lin and Wayne Xin Zhao and Zhewei Wei and Ji-Rong Wen},
  journal={Frontiers of Computer Science},
  year={2024}
}

@article{holmes2023,
  title={Safety Considerations for {AI} in Education},
  author={Holmes, Wayne and Bialik, Maya and Fadel, Charles},
  journal={AI in Education},
  year={2023}
}

@article{zhang2021portrait,
  title={Comprehensive Quality Assessment Based on Digital Portraits: Framework, Indicators, Model, and Application},
  author={Zhang, Zhi and others},
  journal={China Educational Technology},
  number={8},
  pages={25--33},
  year={2021}
}

@article{zhang2022brain,
  title={Ecological Architecture and Application Scenarios of Artificial Intelligence Education Brain},
  author={Zhang, Zhi and Xu, Bingbing},
  journal={Open Education Research},
  number={02},
  pages={64--72},
  year={2022}
}

@article{zhang2023chatgpt,
  title={The Underlying Logic and Possible Paths of {ChatGPT}/Generative {AI} Reshaping Education},
  author={Zhang, Zhi},
  journal={Journal of East China Normal University (Educational Sciences)},
  volume={41},
  number={7},
  pages={131--142},
  year={2023}
}

@article{zhang2017evaluation,
  title={Construction of Multi-Source and Multi-Dimensional Comprehensive Quality Evaluation Model Based on Big Data},
  author={Zhang, Zhi and Qi, Yeguo},
  journal={China Educational Technology},
  number={09},
  pages={69--77},
  year={2017}
}

@article{yu2020portrait,
  title={Construction of Research Learning Student Portrait Based on Visualized Learning Analytics},
  author={Yu, Minghua and Zhang, Zhi and Zhu, Zhiting},
  journal={China Educational Technology},
  number={12},
  pages={36--43},
  year={2020}
}

@article{zhang2021textbook,
  title={Core Concepts and Technical Implementation of Intelligent Digital Textbook System},
  author={Zhang, Zhi and Liu, Dejian and Xu, Bingbing},
  journal={Open Education Research},
  number={01},
  pages={44--54},
  year={2021}
}

@book{bloom1956,
  title={Taxonomy of Educational Objectives: The Classification of Educational Goals},
  author={Bloom, Benjamin S.},
  publisher={David McKay Company},
  address={New York},
  year={1956}
}

@book{vygotsky1978,
  title={Mind in Society: The Development of Higher Psychological Processes},
  author={Vygotsky, Lev S.},
  publisher={Harvard University Press},
  address={Cambridge, MA},
  year={1978}
}

@book{dweck2006,
  title={Mindset: The New Psychology of Success},
  author={Dweck, Carol S.},
  publisher={Random House},
  address={New York},
  year={2006}
}

@article{winne1995,
  title={Self-Regulated Learning},
  author={Winne, Philip H.},
  journal={Journal of Educational Psychology},
  volume={87},
  number={3},
  pages={349--353},
  year={1995}
}

@article{pintrich2000,
  title={The Role of Goal Orientation in Self-Regulated Learning},
  author={Pintrich, Paul R.},
  journal={Handbook of Self-Regulation},
  pages={451--502},
  year={2000}
}

@article{graesser2005,
  title={AutoTutor: An Intelligent Tutoring System with Mixed-Initiative Dialogue},
  author={Graesser, Arthur C. and Chipman, Peter and Haynes, Brian C. and Olney, Andrew},
  journal={IEEE Transactions on Education},
  volume={48},
  number={4},
  pages={612--618},
  year={2005}
}

@article{anderson1995,
  title={Cognitive Tutors: Lessons Learned},
  author={Anderson, John R. and Corbett, Albert T. and Koedinger, Kenneth R. and Pelletier, Ray},
  journal={Journal of the Learning Sciences},
  volume={4},
  number={2},
  pages={167--207},
  year={1995}
}

@article{kasneci2023,
  title={ChatGPT for Good? On Opportunities and Challenges of Large Language Models for Education},
  author={Kasneci, Enkelejda and Sessler, Kathrin and Kuchemann, Sarah and Bannert, Maria and Dementieva, Daryna and Fischer, Frank and Gligorov, Atanas and Heberle, Anne and Heindl, Lena and Kohl, Gregor and others},
  journal={Learning and Individual Differences},
  volume={103},
  pages={102274},
  year={2023}
}

@article{baidoo2023,
  title={Education in the Era of Generative Artificial Intelligence ({AI}): Understanding the Potential Benefits of {ChatGPT} and its Limitations},
  author={Baidoo-Anu, Daniel and Owusu Ansah, Leticia},
  journal={Education and Information Technologies},
  pages={1--16},
  year={2023}
}

@article{luckin2016,
  title={Intelligence Unleashed: An Argument for the Value of {AI} in Education},
  author={Luckin, Rose and Holmes, Wayne and Griffiths, Mark and Forcier, Laurie Burch},
  journal={Pearson Education},
  year={2016}
}

@book{bandura1997,
  title={Self-Efficacy: The Exercise of Control},
  author={Bandura, Albert},
  publisher={W.H. Freeman},
  year={1997}
}

@article{zimmerman2002,
  title={Becoming a Self-Regulated Learner: An Overview},
  author={Zimmerman, Barry J.},
  journal={Theory Into Practice},
  volume={41},
  number={2},
  pages={64--70},
  year={2002}
}

@book{hattie2009,
  title={Visible Learning: A Synthesis of Over 800 Meta-Analyses Relating to Achievement},
  author={Hattie, John},
  publisher={Routledge},
  year={2009}
}

@book{bransford2000,
  title={How People Learn: Brain, Mind, Experience, and School},
  author={Bransford, John D. and Brown, Ann L. and Cocking, Rodney R.},
  publisher={National Academy Press},
  year={2000}
}

@article{socratiq2025,
  title={SocratiQ: A Generative {AI}-Powered Learning Companion for Personalized Education},
  author={Jabbour, Jason and Kleinbard, Kai and Miller, Olivia and Haussman, Robert and Reddi, Vijay Janapa},
  journal={arXiv preprint arXiv:2502.00341},
  year={2025}
}

\end{document}